\documentclass[aip,jcp,reprint,noshowkeys]{revtex4-1}
\usepackage{graphicx,tabularx,dcolumn,bm,xcolor,microtype,multirow,amsmath,amssymb,amsfonts,physics,float,soul,longtable}
\usepackage[version=4]{mhchem}

\usepackage{natbib}
\AtBeginDocument{\nocite{achemso-control}}
\usepackage{mathpazo,libertine}
\usepackage{hyperref}
\hypersetup{
    colorlinks=true,
    linkcolor=blue,
    filecolor=blue,      
    urlcolor=blue,	
    citecolor=blue
}

\newcommand{\mcc}[1]{\multicolumn{1}{c}{#1}}

\newcommand{\Ead}{E^\text{adia}}
\newcommand{\EOO}{E^\text{0-0}}
\newcommand{\Ea}{E_\text{abs}^\text{vert}}
\newcommand{\Ef}{E_\text{fluo}^\text{vert}}
\newcommand{\EreorgES}{E_\text{reorg}^\text{ES}}
\newcommand{\EreorgGS}{E_\text{reorg}^\text{GS}}

\newcommand{\Ezpve}{\Delta E^\text{ZPVE}}

\newcommand{\LCPQ}{Laboratoire de Chimie et Physique Quantiques, Universit\'e de Toulouse, CNRS, UPS, France}
\newcommand{\CEISAM}{Laboratoire CEISAM - UMR CNRS 6230, Universit\'e de Nantes, 2 Rue de la Houssini\`ere, BP 92208, 44322 Nantes Cedex 3, France}

\begin{document}

\title{Evaluating 0-0 Energies with Theoretical Tools: a Short Review}

\author{Pierre-Fran\c{c}ois \surname{Loos}}
\affiliation{\LCPQ}
\author{Denis \surname{Jacquemin}}
\email{Denis.Jacquemin@univ-nantes.fr}
\affiliation{\CEISAM}

\begin{abstract}
For a given electronic excited state, the 0-0 energy ($T_0$ or $T_{00}$) is the simplest property allowing straightforward and physically-sound comparisons between
theory and (accurate) experiment.  However, the computation of 0-0 energies with \emph{ab initio} approaches requires determining both the structure and the vibrational 
frequencies of the excited state, which limits the quality of the theoretical models that can be considered in practice. This explains why only a rather limited, yet constantly
increasing, number of works have been devoted to the determination of this property. In this contribution, we review these efforts with a focus on benchmark studies
carried out for both gas phase and solvated compounds. Over the years, not only as the size of the molecules increased, but the refinement of the theoretical tools has 
followed the same trend. Though the results obtained in these benchmarks significantly depend on both the details of the protocol and the nature of the excited states, 
one can now roughly estimate, in the case of valence transitions, the overall accuracy of theoretical schemes as follows: $1$ eV for CIS, $0.2$--$0.3$ eV for CIS(D), 
$0.2$--$0.4$ eV for TD-DFT when one employs hybrid functionals, $0.1$--$0.2$ eV for ADC(2) and CC2, and $0.04$ eV for CC3, the latter approach being the only one 
delivering chemical accuracy on a near-systematic basis.
\end{abstract}

\maketitle

%
%
\section{Introduction}

Most theoretical works investigating the photophysical or photochemical properties of molecules and materials intend to provide insights supplementing experimental measurements. 
To this end, it is most often necessary to apply first-principle approaches allowing to model electronic excited states (ES). A wide array of such approaches is now available to theoretical
chemists. Probably, the two most prominent ES methods are i) time-dependent density-functional theory (TD-DFT) \cite{Ulr12b} that has been originally proposed by Runge 
and Gross, \cite{Run84} but became very popular under the efficient linear-response (LR) formalism developed by Casida in 1995, \cite{Cas95} and ii) multi-configuration/complete 
active space self-consistent field (MCSCF/CASSCF) theories, \cite{Roo96} that are inherently adapted to model photochemical events.  However, both approaches suffer from significant 
drawbacks. As TD-DFT has been applied for modeling thousands of molecules, the deficiencies of its common adiabatic approximation are now well known, and one can cite important difficulties 
in accurately modeling charge-transfer states, \cite{Toz99b,Dre03,Sob03,Dre04} and Rydberg states, \cite{Toz98,Toz00,Cas98,Cas00} singlet-triplet gaps, \cite{Pea11,Pea12,Pea13,Sun15} 
as well as ES characterized by a significant double excitation character.\cite{Lev06b,Toz00,Ell11} In addition, even for ``well-behaved'' low-lying valence ES, TD-DFT presents a rather significant 
dependency on the exchange-correlation functional (XCF), \cite{Lau13} and choosing an appropriate XCF remains a difficult task.  Similarly, there is also no unambiguous way to select an 
active space in CASSCF calculations, a method, that additionally yields too large transition energies as it does not account for dynamical correlation effects.  Beyond these two very popular theories, 
there exists many alternatives. In the case of single-determinant methods, let us cite i) the Bethe-Salpeter formalism applied on top of the $GW$ approximation (BSE@$GW$), which can be considered 
as a \emph{beyond}-TD-DFT approach and has shown some encouraging performances for chemical systems, \cite{Bla18} ii) the configuration interaction singles with a perturbative double correction 
[CIS(D)], \cite{Hea94,Hea95} the simplest post-Hartree-Fock (HF) method providing reasonably accurate transition energies,  iii) the algebraic diagrammatic construction (ADC) approach, \cite{Dre15} 
whose second-order approximation, ADC(2), enjoys a very favorable accuracy/cost ratio, and iv) coupled cluster (CC) schemes which allow for a systematic theoretical improvement via an increase 
of the expansion order (e.g., comparing CC2, \cite{Chr95}  CCSD, \cite{Koc90,Sta93}  CC3, \cite{Chr95} etc. results), though such strategy comes with a quick inflation of the computational cost.  It is 
also possible to improve CASSCF results by including dynamical correlation effects, typically by applying a second-order perturbative (PT2) correction such as in CASPT2 \cite{And90,And92} or in 
second-order $n$-electron valence state perturbation theory (NEVPT2). \cite{Ang01}  Both theories greatly improve the quality of the transition energies, but become unpractically demanding 
for medium and large systems. Alternatively, one can also compute very high quality transition energies for various types of excited states using selected configuration interaction (sCI) methods 
\cite{Ben69,Whi69,Hur73} which have recently demonstrated their ability to reach near full CI (FCI) quality energies for small molecules. \cite{Hol16,Sha17,Gar17,Gar18,Chi18,Sce18,Loo18a} The idea 
behind such methods is to avoid the exponential increase of the size of the CI expansion by retaining the most energetically relevant determinants only, thanks to the use of a second-order energetic 
criterion to select perturbatively determinants in the FCI space. \cite{Gin13,Gin15}  However, although the \textit{``exponential wall''} is pushed back, this type of methods is only applicable to molecules 
with a small number of heavy atoms with relatively compact basis sets.

\begin{figure}
  \includegraphics[width=\linewidth]{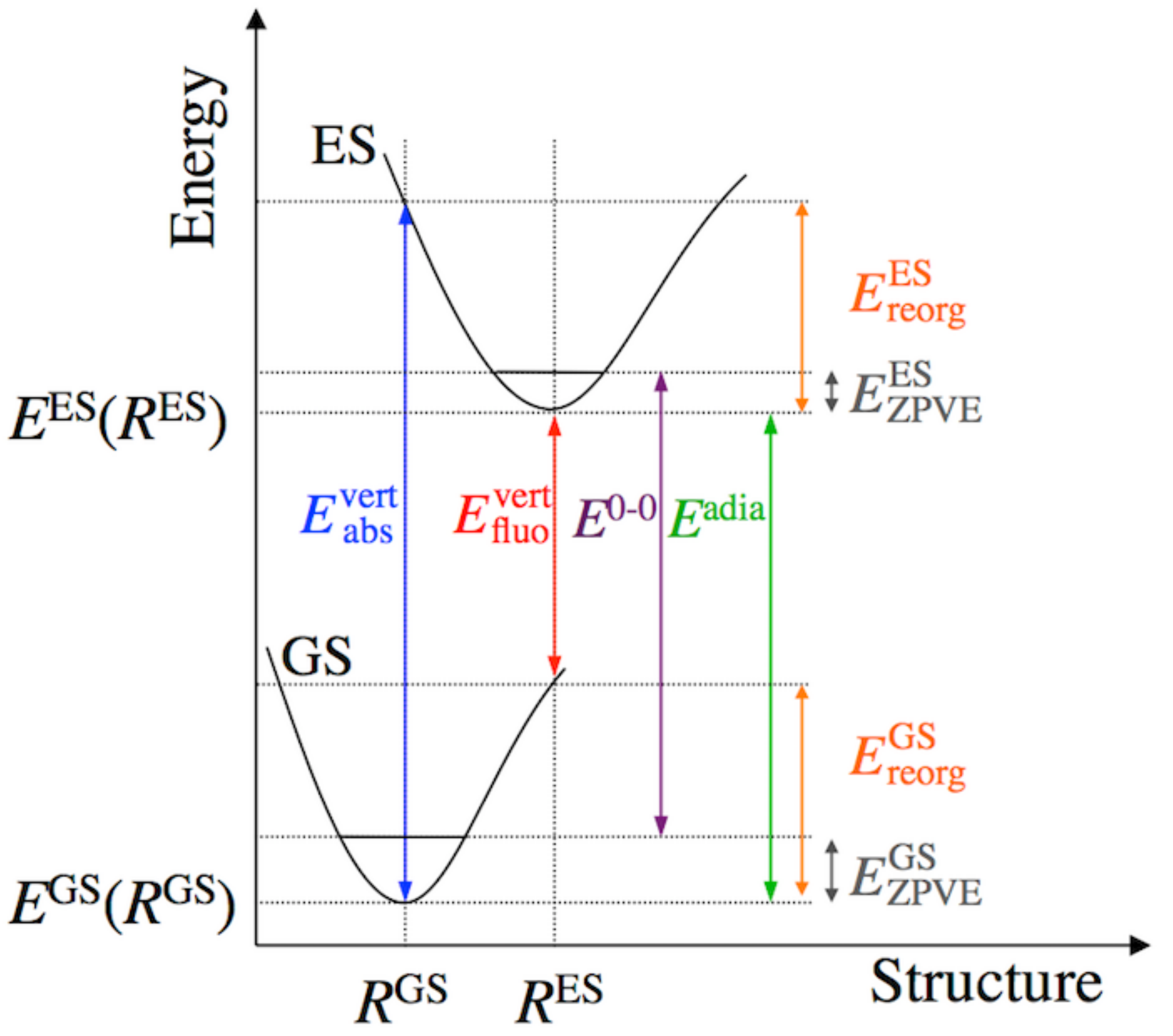}
  \caption{Representation of transition energies between two potential energy surfaces. $\Ea$ (blue) and $\Ef$ (red) are the (vertical) absorption and fluorescence energies, whereas
  $\EreorgGS$ and $\EreorgES$ (orange) are the (geometrical) reorganization energies of the GS and ES states, respectively. $\Ea$ and $\EOO$, our main interests here, are defined in green and purple, respectively. }
   \label{Fig-1}
\end{figure}

Beyond, these important methodological aspects, another issue is that most  \emph{ab initio} calculations of ES properties do not offer direct comparisons with experiment. This is in sharp contrast with 
ground state (GS) properties for which such comparisons are often straightforward. For instance, ``experimental'' ES dipole moments are often determined by indirect procedures, such as the
measurement of solvatofluorochromic effects, so that rather large error bars are not uncommon.  Another example comes with geometries: while there exists an almost infinite number of GS geometries obtained 
through X-ray diffraction techniques for molecules of any size and nature, the experimental determination of ES geometrical parameters remains tortuous, as it typically originates from an analysis of
highly-excited vibronic bands. As a consequence, experimental ES structures are available only for a handful of small compounds, prohibiting comparisons between theory and experiment for
non-trivial structures.  Although, for both ES dipole moments and geometries, theoretical approaches have therefore a clear edge over their experimental counterparts, such calculations nevertheless 
require the access to ES energy gradients, which limits the number of methods that can be applied for non-trivial compounds. Besides, the most commonly reported theoretical ES data, that is, vertical 
absorption energies, have no experimental counterpart as they correspond to vibrationless differences between total ES and GS energies at the GS geometry ($\Ea$ in Figure \ref{Fig-1}).
As a consequence, they can be used to compare trends in a homologous series of compounds, \cite{Lau14} but are rather useless when one aims for quantitative theory-experiment 
comparisons.  Therefore, the simplest ES properties that are well-defined both theoretically and experimentally are the 0-0 energies ($\EOO$, sometimes denoted $T_0$ or $T_{00}$). For a 
given ES, the 0-0 energy corresponds to the difference  between the ES and GS energies at their respective geometrical minimum, the adiabatic energy $\Ead$ (sometimes denoted $T_e$), 
corrected by the difference of zero-point vibrational energies between these two states  ($\Ezpve$).  For gas phase molecules with well-resolved vibronic spectrum, $\EOO$ can be directly measured with 
uncertainties of the order of 1 cm$^{-1}$. In other words, extremely accurate experimental data are available. In solution, $\EOO$ is generally defined as the crossing point between the measured (normalized) 
absorption and emission spectra. On the theory side, whilst $\EOO$ is a well defined quantity, its calculation is no cakewalk, notably due to the $\Ezpve$ term that necessitates the estimation of the 
vibrational ES frequencies.

In the present mini-review, we will consider previous works dealing with theory-experiment comparisons for $\Ead$ or $\EOO$ energies. As expected, over the years, the methods available to compute $\EOO$
have dramatically improved, so as the accuracy. Here, we do focus on benchmark studies tackling a significant number of diverse molecules with first principle methods. We do 
not intend to provide an exhaustive list of the works considering only one or two compounds and their comparison with experiment, or a specific chemical family of compounds. For the second category, the
interested reader can find several works devoted to, e.g., fluoroborate derivatives, \cite{Chi13a,Chi13b,Chi14c} biological chromophores, \cite{Kam11,Upp12b} DNA bases, \cite{Ovc14} cyanines, \cite{Aza18}
coumarins, \cite{Mun15} as well as many other works focussed on band shapes rather than $\EOO$ energies. \cite{San07a,Pet07,San08b,Avi13,Bai13,Avi13b,Bar14,Bar14e,San16b} 

\section{0-0 energies computed in gas phase}

In this Section, we review the theoretical investigations relying on gas-phase calculations to obtain $\Ead$ or $\EOO$. Though there is no universal classification for molecule sizes, we first discuss works
focussing on small compounds, that is, sets of compounds largely dominated by di- and tri-atomic molecules, before turning to medium (e.g., benzene) and large (e.g., real-life
dyes) molecules in the second subsection.  The main information associated with the various studies discussed below are summarized in Table \ref{Table-1}.

%
%
\begin{longtable*}{lcclldd}
\caption{
\label{Table-1}
Statistical analysis of the results obtained in various benchmarks comparing gas-phase $\Ead$ or $\EOO$ computations to experimental data. MSE and MAE are the mean signed and mean absolute
errors, and are given in eV. When a different method was used to compute $\Ead$ and to obtain the structures (and ZPVE corrections) this is mentioned using the usual ``//'' notation. 
}
\\
\hline
\hline
Ref.				&	Year		& No.~of ESs	& No.~of molecules		& Method 				&	\mcc{MSE}		&	\mcc{MAE}		\\
\hline
\endfirsthead
\hline
\hline
Ref.				&	Year		& No.~of ESs	& No.~of molecules		& Method 				&	\mcc{MSE}		&	\mcc{MAE}		\\
\hline
\endhead
\hline \multicolumn{7}{r}{Continued on next page} \\
\endfoot
\hline
\hline
 \multicolumn{7}{l}{$^a${$\Ead$ values were considered;}}\\
 \multicolumn{7}{l}{$^b${Depending on the molecule $\Ead$ or $\EOO$ values were considered;}}\\
 \multicolumn{7}{l}{$^c${Some of the experiments were made in solution or in a matrix, but the the gas-phase theoretical calculations were uncorrected;}}\\
 \multicolumn{7}{l}{$^d${Solvent effects empirically corrected;}}\\
 \multicolumn{7}{l}{$^e${Same set (GI) as in Ref.~\citenum{Gri04b};}}\\
 \multicolumn{7}{l}{$^f${Same set (KH) as in Ref.~\citenum{Koh03};}}\\
  \multicolumn{7}{l}{$^g${$\Ezpve$ at the B3LYP level;}}\\
  \multicolumn{7}{l}{$^h${(Sub)set (SKF) of the one considered in Ref.~\citenum{Sen11b};}}\\
 \multicolumn{7}{l}{$^i${More than one conformer of the same molecules are investigated in several cases;}}\\
 \multicolumn{7}{l}{$^j${Same set (WGLH) as in Ref.~\citenum{Win13};}}\\
 \multicolumn{7}{l}{$^k${Variant ``A'' of the spin-scaling parameters, the so-called ``original'' values;}}\\
\endlastfoot
\citenum{Sta95}$^a$	&	1995		&	6		&6 (diatomics)			& CIS/aug-cc-pVTZ		&	-0.06		&	0.73		\\
				&			&			&					& CIS(D)/aug-cc-pVTZ	&	 0.27		&	0.27		\\
				&			&			&					& CCSD/aug-cc-pVTZ	&  	-0.19		&	0.19		\\
\citenum{Fur02}$^b$	&	2002		& 	34		&28 (mostly di/triatomics)	& CIS/aug-TZVPP		&	0.32		&	0.66		\\
				&			&			&					& TD-HF/aug-TZVPP	&	0.23		&	0.63		\\
				&			&			&					& LDA/aug-TZVPP		&	-0.18		&	0.25		\\
				&			&			&					& BLYP/aug-TZVPP		&	-0.27		&	0.32		\\
				&			&			&					& BP86aug-TZVPP		&	-0.22		&	0.31		\\
				&			&			&					& PBE/aug-TZVPP		&	-0.24		&	0.30		\\
				&			&			&					& B3LYP/aug-TZVPP	&	-0.13		&	0.28		\\
				&			&			&					& PBE0/aug-TZVPP		&	-0.08		&	0.30		\\
\citenum{Koh03}$^b$&	2003		&	20		& 29  (mostly di/triatomics)& CC2/aug-cc-pVQZ		&	-0.05		&	0.17		\\
\citenum{Die04}$^c$	&	2004		&	9		& 7 (aromatics)			& B3LYP/TZVP			&	-0.13		&	0.43		\\
\citenum{Die04b}$^d$&	2004		&	43		& 41 ($\pi$-conjugated)	& BP86/TZVP			&	-0.56		&	0.57		\\
				&			&			&					& B3LYP/TZVP			&	-0.33		&	0.35		\\
				&			&			&					& BHHLYP/TZVP		&	-0.01		&	0.18		\\
\citenum{Gri04b}	&	2004		&	32		& 22	(diverse)			& B3LYP/TZV(d,p)		&	-0.11		&	0.28		\\
				&			&			&	& CIS(D)/aug-cc-pVTZ//B3LYP/TZV(d,p)		&	0.16		&	0.19		\\
				&			&			&	& SCS-CIS(D)/aug-cc-pVTZ//B3LYP/TZV(d,p)	&	0.23		&	0.23		\\
\citenum{Hat05c}$^a$&	2005		&	19		& 4 (diatomics)			& CIS/aug-cc-pwCVQZ 	&	0.03		&	0.57		\\
				&			&			&					& CIS(D)/aug-cc-pwCVQZ &	0.29		&	0.26		\\
				&			&			&					& ADC(2)/aug-cc-pwCVQZ&	0.18		&	0.21		\\
				&			&			&					& CC2/aug-cc-pwCVQZ 	&	0.10		&	0.16		\\
				&			&			&					& CCSD/aug-cc-pwCVQZ	&	0.20		&	0.20		\\
				&			&			&				& CCSDR(3)/aug-cc-pwCVQZ	&	0.07		&	0.07		\\	
				&			&			&					& CC3/aug-cc-pwCVQZ	&	0.01		&	0.04		\\
\citenum{Rhe07}	&	2007		&	32		& 22	(diverse)$^e$& CIS/aug-cc-pVTZ//CIS/6-311G(d,p)&	0.63		&	0.71		\\
				&			&			&		& CIS(D)/aug-cc-pVTZ//CIS/6-311G(d,p)	& 	0.19		&	0.22		\\	
				&			&			&	& SCS-CIS(D)/aug-cc-pVTZ//CIS/6-311G(d,p)	& 	0.02		&	0.12		\\		
				&			&			&	& SOS-CIS(D)/aug-cc-pVTZ//CIS/6-311G(d,p)	& 	0.02		&	0.12		\\	
\citenum{Hel08}$^a$	&	2008		&	26		&	19 (di/triatomics)	&CC2/cc-pVQZ			&	0.01		&	0.17		\\
				&			&			&					&SCS-CC2/cc-pVQZ		&	0.09		&	0.16		\\
				&			&			&					&SOS-CC2/cc-pVQZ		&	0.13		&	0.17		\\	
				&			&	32	& 22	(diverse)$^e$&B3LYP/aug-cc-pVTZ//B3LYP/TZVP	&	-0.13		&	0.29		\\
				&			&			&			& CC2/aug-cc-pVTZ//B3LYP/TZVP	&	-0.02		&	0.14		\\
				&			&			&		&SCS-CC2/aug-cc-pVTZ//B3LYP/TZVP	&	0.08		&	0.14		\\
				&			&			&		&SOS-CC2/aug-cc-pVTZ//B3LYP/TZVP	&	0.13		&	0.17		\\
\citenum{Rhe09}$^b$&	2009		&	20	& 29 (mostly di/triatomics)$^f$	& CIS/aug-cc-pVTZ		&	0.19		&	0.58		\\
				&			&			&			& SOS-CIS(D$_0$)/aug-cc-pVTZ	&	0.12		&	0.26		\\
				&			&			&					& CC2/aug-cc-pVTZ		&	-0.08		&	0.18		\\
				&			&	 32		& 22	(diverse)$^e$& SOS-CIS(D$_0$)/aug-cc-pVTZ	&	0.05		& 	0.17		\\
\citenum{Liu10}$^b$&	2010		&	20	& 29 (mostly di/triatomics)$^f$& B3LYP/aug-cc-pVTZ	& 	-0.25		&	0.30		\\
				&			&			&				& TDA-B3LYP/aug-cc-pVTZ	&	-0.12		&	0.26		\\
				&			&			&				& $\omega$B97/aug-cc-pVTZ	&	-0.05		&	0.25		\\
				&			&			&			& TDA- $\omega$B97/aug-cc-pVTZ	&	0.08		&	0.25		\\
\citenum{Ngu10}$^g$&	2010		&	9		& 7 (charge-transfer)		& B3LYP/6-311+G(d,p)	&	-0.36		&	0.36		\\
				&			&			&					& LC-BOP/6-311+G(d,p)	&	0.16		&	0.24		\\
				&			&			&				& CAM-B3LYP/6-311+G(d,p)	&	0.07		&	0.22		\\
				&			&			&				& MCAM-B3LYP/6-311+G(d,p)	&	-0.06		&	0.06		\\
\citenum{Sen11b}	&	2011		&	91		& 109 (diverse)			& CIS/def2-TZVP//B3LYP/def2-TZVP		&	0.90		&	0.98		\\
				&			&			&					& LSDA/def2-TZVP//B3LYP/def2-TZVP		&	-0.21		&	0.49		\\
				&			&			&					& PBE/def2-TZVP//B3LYP/def2-TZVP		&	-0.33		&	0.40		\\
				&			&			&					& BP86/def2-TZVP//B3LYP/def2-TZVP		&	-0.32		&	0.39		\\
				&			&			&					& TPSS/def2-TZVP//B3LYP/def2-TZVP		&	-0.20		&	0.32		\\
				&			&			&					& B3LYP/def2-TZVP						&	-0.08		&	0.21		\\
				&			&			&					& PBE0/def2-TZVP//B3LYP/def2-TZVP		&	-0.08		&	0.25		\\
				&			&	15		& 15 (subset of previous)	& CC2/def2-TZVPD//B3LYP/def2-TZVP		&	0.10		&	0.17		\\
\citenum{Bat12}	&	2012		&	91		& 109 (various)$^h$		& cTPSS/def2-TZVP//B3LYP/def2-TZVP		&	-0.26		&	0.34		\\
				&			&			&					& TPSSh/def2-TZVP//B3LYP/def2-TZVP		&	-0.08		&	0.26		\\
				&			&			&					& cTPPSh/def2-TZVP//B3LYP/def2-TZVP		&	-0.13		&	0.27		\\
\citenum{Win13}	&	2013		&	66		& 46 (aromatics)$^i$		& B3LYP/aug-cc-pVTZ//B3LYP/def2-TZVP	&	0.00		&	0.19		\\	
				&			&			&					& ADC(2)/aug-cc-pVTZ//ADC(2)/def2-TZVPP	&	-0.03		&	0.08		\\
				&			&			&					& CC2/aug-cc-pVTZ//CC2/def2-TZVPP		&	0.00		&	0.07		\\
				&			&			&				& SCS-CC2/aug-cc-pVTZ//SCS-CC2/def2-TZVPP	&	0.01		&	0.05		\\
				&			&			&				& SOS-CC2/aug-cc-pVTZ//SOS-CC2/def2-TZVPP	&	-0.01		&	0.06		\\
\citenum{Fan14b}	&	2014		&	79		& 96 (various)$^h$		& CIS/cc-pVDZ							&	0.78		&	0.88		\\	
				&			&			&					& CC2/cc-pVDZ						&	0.11		&	0.19		\\
				&			&			&					& BP86/cc-pVDZ						&	-0.38		&	0.42		\\
				&			&			&					& B3LYP/cc-pVDZ						&	-0.11		&	0.24		\\
				&			&			&					& PBE0/cc-pVDZ						&	-0.03		&	0.26		\\
				&			&			&					& M06-2X/cc-pVDZ						&	0.05		&	0.30		\\
				&			&			&					& M06-HF/cc-pVDZ						&	-0.01		&	0.50		\\
				&			&			&					& CAM-B3LYP/cc-pVDZ					&	0.09		&	0.27		\\	
				&			&			&					& $\omega$B97X-D/cc-pVDZ				&	0.10		&	0.27		\\
\citenum{Bar14b}	&	2014		&	29		& 15 (small radicals)		&CIS/6-311++G(d,p)						&	1.66		&	1.75		\\
				&			&			&					&BLYP/6-311++G(d,p)					&	-0.22		&	0.32		\\
				&			&			&					&PBE/6-311++G(d,p)					&	-0.13		&	0.29		\\
				&			&			&					&VSXC/6-311++G(d,p)					&	-0.07		&	0.26		\\
				&			&			&					&M06-L/6-311++G(d,p)					&	0.17		&	0.36		\\
				&			&			&					&B3LYP/6-311++G(d,p)					&	-0.05		&	0.18		\\
				&			&			&					&PBE0/6-311++G(d,p)					&	0.05		&	0.25		\\
				&			&			&					&M06/6-311++G(d,p)					&	-0.10		&	0.25		\\
				&			&			&					&BHandH/6-311++G(d,p)					&	0.16		&	0.32		\\
				&			&			&					&BHandHLYP/6-311++G(d,p)				&	0.11		&	0.35		\\
				&			&			&					&M06-2X/6-311++G(d,p)					&	-0.04		&	0.24		\\
				&			&			&					&CAM-B3LYP/6-311++G(d,p)				&	0.08		&	0.23		\\
				&			&			&					&$\omega$B97X-D/6-311++G(d,p)			&	0.08		&	0.22		\\
				&			&			&					&LC-BLYP/6-311++G(d,p)					&	0.18		&	0.38		\\
				&			&			&					&LC-PBE/6-311++G(d,p)					&	0.28		&	0.45		\\
				&			&			&					&LC-M06-L/6-311++G(d,p)				&	0.33		&	0.39		\\
				&			&			&					&HSE06/6-311++G(d,p)					&	0.08		&	0.22		\\
				&			&			&					&HISS/6-311++G(d,p)					&	0.29		&	0.38		\\
				&			&			&					&CASPT2/6-311++G(d,p)					&	-0.02		&	0.12		\\
\citenum{Tun16}$^g$&	2016		& 	68		& 59 (organic)$^h$		& OM2/MRCI							&	-0.01		&	0.26		\\
				&			&			&					& OM3/MRCI			 				&	-0.03		&	0.27		\\
				&			&			&					&  B3LYP/def2-TZVP					& 	-0.11		&	0.24		\\			
				&			&	65		& 45 (aromatics)$^j$	&OM2/MRCI								&	-0.22		&	0.35		\\
				&			&			&					& OM3/MRCI			 				&	-0.23		&	0.35		\\
\citenum{Oru16}	&	2016		&	66		& 46 (aromatics)$^j$	& CIS/def2-TZVP						&	1.08		&	1.08		\\
				&			&			&					& BP86/def2-TZVP						&	-0.39		&	0.40		\\
				&			&			&					& B3LYP/def2-TZVP						&	0.05		&	0.20		\\
				&			&			&					& PBE0/def2-TZVP						&	0.16		&	0.24		\\
				&			&			&					& M06-2X/def2-TZVP					&	0.33		&	0.36		\\
				&			&			&					& M06-HF/def2-TZVP					&	0.55		&	0.57		\\
				&			&			&					& CAM-B3LYP/def2-TZVP					&	0.30		&	0.33		\\
				&			&			&					&  $\omega$B97X-D/def2-TZVP			&	0.30		&	0.32		\\
				&			&			&					&  CC2/def2-TZVP						&	0.09		&	0.11		\\
\citenum{Sch17}$^k$&	2017		&	66		& 46 (aromatics)$^j$	& B2PLYP/aug-cc-pVTZ//B3LYP/def2-TZVP	&	0.01		&	0.11		\\
				&			&			&					& B2GPPLYP/aug-cc-pVTZ//B3LYP/def2-TZVP	&	0.21		&	0.24		\\
				&			&			&					& DSD-BLYP/aug-cc-pVTZ//B3LYP/def2-TZVP	&	0.05		&	0.10		\\
				&			&			&				& DSD-PBEP86/aug-cc-pVTZ//B3LYP/def2-TZVP	&	0.03		&	0.08		\\
				&			&			&					& PBE0-2/aug-cc-pVTZ//B3LYP/def2-TZVP	&	0.19		&	0.21		\\
				&			&			&					& PBE0-DH/aug-cc-pVTZ//B3LYP/def2-TZVP	&	0.25		&	0.28		\\
				&			&			&				& B2PLYP/aug-cc-pVTZ//SCS-CC2/def2-TZVPP	&	-0.01		&	0.10		\\
				&			&			&				& B2GPPLYP/aug-cc-pVTZ//SCS-CC2/def2-TZVPP	&	0.07		&	0.10		\\
				&			&			&				& DSD-BLYP/aug-cc-pVTZ//SCS-CC2/def2-TZVPP	&	0.02		&	0.06		\\
				&			&			&			& DSD-PBEP86/aug-cc-pVTZ//SCS-CC2/def2-TZVPP	&	-0.02		&	0.06		\\
				&			&			&				& PBE0-2/aug-cc-pVTZ//SCS-CC2/def2-TZVPP	&	0.15		&	0.17		\\
				&			&			&				& PBE0-DH/aug-cc-pVTZ//SCS-CC2/def2-TZVPP	&	0.25		&	0.28		\\
\citenum{Loo18b}$^g$&	2018		&	35		& 31 (medium-size organic)	& CC3/aug-cc-pVTZ//CCSDR(3)/def2-TZVPP	&-0.01	&	0.02		\\
				&			&			&				& CCSDR(3)/aug-cc-pVTZ//CCSDR(3)/def2-TZVPP	&	0.04		&	0.05		\\
				&			&			&					& CCSD/aug-cc-pVTZ//CCSDR(3)/def2-TZVPP	&	0.21		& 	0.21		\\
				&			&			&					& CC2/aug-cc-pVTZ//CCSDR(3)/def2-TZVPP	&	0.04		& 	0.08		\\
\citenum{Loo19a}$^g$&	2019		&	119		& 109 (diverse)			& CC3/aug-cc-pVTZ//CCSD/def2-TZVPP		&	-0.01		&	0.03		\\
\end{longtable*}

\subsection{Small compounds}

To the best of our knowledge, one of the first investigation on adiabatic energies is due to Stanton and coworkers, \cite{Sta95} who compared the performances of CIS, CIS(D), and CCSD for the computation of
$\Ead$ in six diatomic molecules (\ce{H2}, \ce{BH}, \ce{CO}, \ce{N2}, \ce{BF}, and \ce{C2}) in 1995. For such small molecules, it is possible to analyze the spectroscopic data \cite{Hub79} to obtain directly experimental
$\Ead$ rather than $\EOO$. \cite{Odd85}  Three atomic basis set were considered, namely, 6-31G(d), aug-cc-pVDZ, and aug-cc-pVTZ; we report only the results obtained with the largest basis in Table \ref{Table-1}.
It is crystal clear that the CIS method is very far from experiment even for these quite simple molecules, with errors ranging from $+0.99$ eV (\ce{N2}) to $-2.34$ eV (\ce{C2}). The inclusion of
the perturbative doubles vastly improves the estimates with a mean absolute error (MAE) of $0.27$ eV. Nonetheless, CIS(D) systematically overshoots the experimental values for this particular set.
CCSD further reduces the absolute error but underestimates $\Ead$ in each case. We note that such error sign is rather unusual for CCSD. Indeed, this approach generally delivers, for valence ES, too 
large transition energies. \cite{Sch08,Sil08} The trend obtained in this early study is therefore most probably related to the size of the considered molecules. \cite{Loo18a}

A second key investigation is due to Furche and Alrichs (FA), \cite{Fur02,Rap05} who benefitted from pioneering developments and efficient implementation of TD-DFT energy gradients. \cite{Cai99}  Using this approach,
they investigated around thirty small-size compounds (except for glyoxal, pyridine, benzene, and porphyrin) using a quite large basis set and several XCF. As can be seen in Table \ref{Table-1},
the two HF-based approaches, CIS and TD-HF, deliver very large errors, with a positive MSE, as expected for methods neglecting dynamical correlation.  All the XCF tested within TD-DFT give a MAE 
in the $0.25$--$0.32$ eV range, with no clear-cut advantage for hybrids over semi-local functionals, an outcome probably related to the size of the molecules.  Small subsets of the original FA 
set were considered by Chiba \emph{et al.}, \cite{Chi06} and Nguyen \emph{et al.} \cite{Ngu10} for the testing of their own implementations of TD-DFT gradients for range-separated hybrids
(not shown in Table \ref{Table-1}). In 2003, K\"ohn and H\"attig (KH) estimated transition energies for a similar set as FA with their own implementation of CC2 gradients. \cite{Koh03}  
These authors considered several atomic basis sets and we report in Table \ref{Table-1} the data computed with the quadruple-$\zeta$ basis, though the deviations with respect to 
the triple-$\zeta$ basis are rather insignificant. As can be seen the CC2 MAE ($0.17$ eV) is significantly smaller than its TD-DFT counterparts. For a work carried out more than 15 years ago, it is remarkable 
that a CC2 estimate of $\EOO$ could be computed for a quite large molecule such as azobenzene. The KH set was employed twice in the following years. First, by Rhee, Casanova, 
and Head-Gordon in 2009 when they proposed the SOS-CIS(D$_0$) method which gives a MAE of $0.26$ eV. \cite{Rhe09} Second, by Liu \emph{et al.} in 2010, who found that both TD-DFT and its 
Tamm-Dancoff approximation (TDA) deliver similar average deviations while considering B3LYP and $\omega$B97 as XCF. Indeed, the differences between the TD-DFT and TDA results 
(average errors of $0.12$ and $0.14$ eV with B3LYP and $\omega$B97, respectively) are significantly smaller than the discrepancies with respect to experiment. In addition, H\"attig's group also considers a 
similar set of compounds in 2008 to investigate spin-scaled variants of CC2. They found that the average deviations were not significantly altered compared to conventional CC2, and that the spin-scaling 
version improved the overall consistency (correlation) compared to experiment. \cite{Hel08}

In 2005, H\"{a}ttig evaluated the performances of various single-reference wavefunction approaches using 19 ES (11 singlet and 8 triplet) determined on four diatomic molecules (\ce{N2}, \ce{CO}, \ce{CF}, and \ce{BH}) using 
a huge basis set allowing to be near the complete basis set limit. \cite{Hat05c} As can be deduced from Table \ref{Table-1}, the convergence with respect to the expansion order in the CC series (CIS, CC2, 
CCSD, CCSDR(3), CC3) is rather erratic. In addition, all approaches (partially) including contributions from the doubles, i.e., CIS(D), ADC(2), CC2, and CCSD provide similar results with MAE of ca.~$0.2$ eV. 
In contrast, the inclusion of triples, either perturbatively or iteratively, leads to average deviations smaller than $0.10$ eV. To our knowledge, this work was the first demonstration that ``chemical accurate'' $\Ead$
(errors smaller than $1$ kcal/mol or $0.043$ eV) could potentially be attained with theoretical methods on an almost systematic basis.

\subsection{Medium and large compounds}

The first studies considering the computation of $\EOO$ in larger, ``real-life'' structures are due to Grimme and his collaborators in 2004. \cite{Die04,Die04b,Gri04b} In the first work of their series, \cite{Die04} they investigated the
vibronic shapes of seven $\pi$-conjugated molecules (anthracene, azulene, octatretraene, pentacene, phenoxyl radical, pyrene, and styrene) with TD-B3LYP. The reproduction of the experimental
band shapes is generally excellent, but the error in $\EOO$ compared to experiment (ranging from $-0.69$ eV to $+0.86$ eV) is rather large, leading to the conclusion that
the quality of the TD-DFT transition energies have to be blame rather than the structures, at least, for these rigid aromatic molecules.  \cite{Die04} In their second paper, \cite{Die04b} the number of transitions was significantly increased as they
studied 30  singlet-singlet transitions and 13 doublet-doublet transitions in $\pi$-conjugated compounds.  The calculations were performed with TD-DFT in gas-phase with three XCF (BP86, B3LYP,
and BHHLYP) and the solvent effects were accounted  by applying an empirical $+0.15$ eV shift to the experimental 0-0 energies measured in condensed phase. Dierksen and Grimme noted a smooth evolution 
of the computed $\EOO$ energies with the amount of \emph{exact} exchange included in the functional for the $\pi \rightarrow \pi^\star$ singlet-singlet transitions, BHHLYP leading to the smallest MAE. \cite{Die04b}  Eventually,
in Ref.~\citenum{Gri04b}, a third test set including 20  $\pi \rightarrow \pi^\star$ and 12  $n \rightarrow \pi^\star$ transitions, the GI set, was designed to compare the performances of TD-DFT, CIS(D), and one of its 
spin-scaled variant, namely SCS-CIS(D). For this set, the CIS(D) approach clearly outperforms TD-B3LYP, whereas SCS-CIS(D) does not improve the overall MAE but delivers a
more balanced description of the two families of ES. Indeed, CIS(D) yields a significantly smaller MAE ($0.10$ eV) for the  $n \rightarrow \pi^\star$  subset than for its  $\pi \rightarrow \pi^\star$ 
counterpart ($0.25$ eV). The GI set was also used in 2008 to evaluate the performances of several CC2 variants which all provided MAE around $0.15$ eV. \cite{Hel08} Though most
wavefunction calculations were performed on TD-DFT geometries, Hellweg \emph{et al.}~also tested the impact of performing CC2 optimizations.  Interestingly, they noted almost no major
difference for the  $\pi \rightarrow \pi^\star$  states, whereas for the  $n \rightarrow \pi^\star$  transitions, CC2 structures significantly redshifted the excitation energies as compared to
those obtained with TD-DFT geometries.  The GI set was also used twice by Head-Gordon and coworkers. \cite{Rhe07,Rhe09} to  evaluate the performances of spin-scaled variants of the 
CIS(D) approach. In their first work, the calculations were made on CIS structures, and the SCS-CIS(D) and SOS-CIS(D) approaches both exhibit very good performances (MAE for both approaches 
0.12 eV), a result probably partially due to error compensations.  \cite{Rhe07} In the second work the focus was set on the performances of SOS-CIS(D$_0$). \cite{Rhe09} In the most refined calculations, 
a double-$\zeta$ basis set was applied to obtain the geometries and ZPVE corrections, whereas $\Ead$ was determined with aug-cc-pVTZ. The accuracy of SOS-CIS(D$_0$) is significantly 
better for the GI set (containing medium-sized compounds) than for the KH set (gathering di/tri-atomics), indicating that the size of the molecules has a significant influence on the methodological conclusions. 
In addition, the MSE for the $\pi \rightarrow \pi^\star$  ($+0.14$ eV) and $n \rightarrow \pi^\star$ ($-0.11$ eV) subsets differ with SOS-CIS(D$_0$), further stressing that reaching a balanced description of ES 
of different natures is difficult.

A decade ago, Nguyen, Day and Pachter compared  TD-DFT/6-311+G(d,p) and experimental adiabatic energies for seven substituted coumarins and two stilbene derivatives exhibiting transitions
with a significant charge-transfer character. \cite{Ngu10} Unsurprisingly, \cite{Dre04,Pea06} range-separated hybrids clearly deliver more accurate results in this set, the B3LYP $\EOO$ being 
systematically too small.

In 2011, Furche's group came up with another popular set (SKF) of 109 $\EOO$ energies obtained in 91 very diverse compounds encompassing small, medium, and large structures for which experimental
gas-phase $\EOO$ values are available. Special care was taken in order to include diverse compounds (organic/inorganic, aliphatic/aromatic, etc.) and ES (86 singlets, 12 triplets, and 11
spin-unrestricted transitons). \cite{Sen11b}  The majority of the results were obtained on B3LYP/def2-TZVP structures and $\Ezpve$, using $\Ead$ determined with various XCF and the same def2-TZVP basis set. 
As detailed below, several protocols were tested.  For this diverse set, there is a significant superiority of the hybrid XCF (B3LYP and PBE0) compared  to the local and semi-local XCF (Table \ref{Table-1}) 
which contrasts with the FA set (containing smaller compounds) discussed above. In Ref. \citenum{Sen11b}, the authors also show that using a (non-augmented) polarized triple-$\zeta$ basis provides 
$\EOO$ within ca.~$0.03$ eV of the basis set limit at the TD-DFT level and that, consistently with Grimme's conclusions, the error on the transition energies must be blame for the the major part of this deviation, 
the variations of the structural parameters when changing XCF having a minor impact. From this larger set, Furche and coworkers also extracted a subset of 15 representative ES, and performed ADC(2) and 
CC2 calculations. These two methods were found to behave similarly and the addition of diffuse functions was found mandatory (in contrast to TD-DFT). For this subset, the MAE is $0.17$ eV with CC2, a 
value consistent with the CC2 MAE obtained for previously discussed sets. A year later, the same group extended their analysis to variants of the TPSS XCF.\cite{Bat12}  They found that the current-dependent formalism 
for TPSS and TPSSh (cTPSS and cTPSSh) yield larger deviations than the standard formalism.  In 2014, Fang, Oruganti, and Durbeej considered a larger number of XCF on a set encompassing all the singlet 
and triplet transitions of the SKF set. \cite{Fan14b}  Overall the most accurate results are attained with CC2, whereas the ``standard'' global and range-separated hybrids (B3LYP, PBE0, CAM-B3LYP and 
$\omega$B97X-D) yield errors around $0.25$ eV. Unsurprisingly CIS and XCF including 100\% of \emph{exact} exchange (M06-HF) overestimate substantially the experimental reference, whereas BP86 gives the 
opposite error sign. In addition, the authors investigated the errors in 9 {chemically-intuitive} subsets. For the organic compounds, CC2 was systematically found to outperform TD-DFT in terms of average error, 
whereas this does not hold for small inorganic compounds.  In an effort to come up with a computationally effective protocol, the authors also studied methodological effects on two quantities. First, 
$\Delta \EOO = \EOO - \Ead$, that is the $\Ezpve$ correction, which was found to be centered on $-0.12$ eV, with a very small methodological dependence: the standard deviations determined across the 
various tested methods was as small as $0.02$ eV, and in the $0.01$--$0.05$ eV range for the nine subsets. This clearly indicates that $\Ezpve$ is rather insensitive to the level of theory, confirming previous 
studies performed in the same research group, \cite{Upp12b} and others. \cite{Jac12d} Second, they studied  $\Delta \Ead = \Ead - \Ea$, that is, the ES reorganization energy, $\EreorgES$. The methodological 
standard deviation was only $0.10$ eV for $\EreorgES$, as compared to the much larger spread for $\Ea$ ($0.39$ eV), indicating that $\EreorgES$ is also much less dependent on the level of theory than the vertical 
energies, in line with previous observations (see above). \cite{Die04} Nevertheless, in contrast to $\Ezpve$, the $\EreorgES$ values cover a broad range of values depending on the molecule ($-0.37$ $\pm$ $0.30$ eV).  
Later, Furche's 2011 set was also selected to assess semi-empirical approaches (see below for details). \cite{Tun16}

\begin{figure}
  \includegraphics[width=\linewidth]{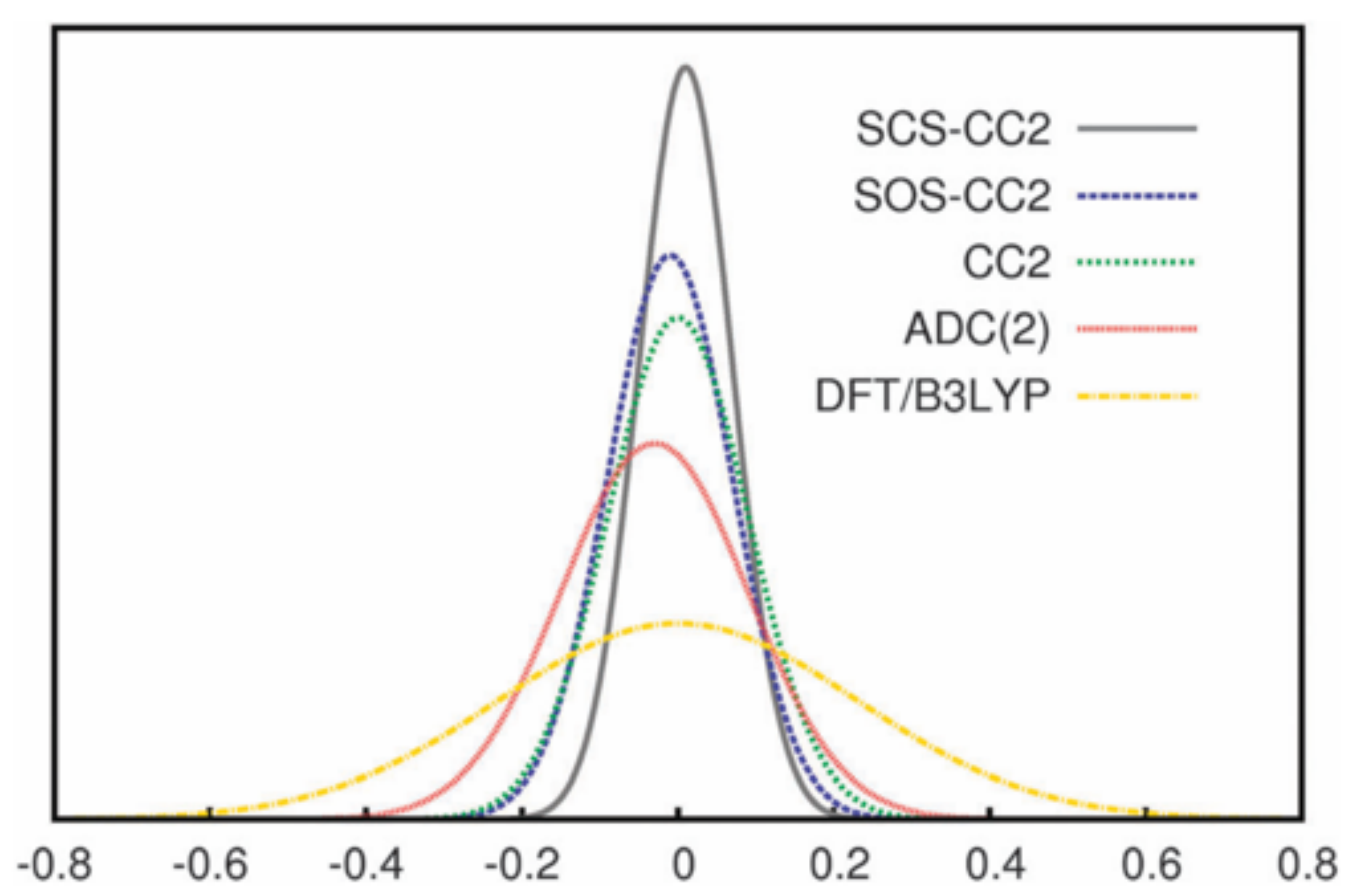}
  \caption{Error distribution pattern for $\EOO$ in the WGHL set of compounds. The values are in eV. Reproduced from Ref.~\citenum{Win13} with permission from the PCCP owner societies.}
   \label{Fig-2}
\end{figure}

Two years later, H\"attig and collaborators compared theoretical $\EOO$ values to highly accurate gas phase experimental references for a 66-singlet set strongly dominated by $\pi \rightarrow \pi^\star$ transitions (63 out of 66) 
in aromatic  organic molecules (substituted phenyls and larger compounds) leading to the WGLH set. \cite{Win13} They rely on the aug-cc-pVTZ basis set for determining $\Ead$, and the def2-TZVPP basis set
for obtaining structures and vibrations. As can be seen in Figure \ref{Fig-2}, second-order wavefunction approaches, i.e., ADC(2), CC2, SCS-CC2, and SOS-CC2 performed beautifully with a tight distribution 
around the experimental reference and very small average deviations, all below the $0.10$ eV threshold. This success is probably partially related to the rather uniform nature of the ES considered in this particular study, as compared to the 
SKF set. Obviously, TD-B3LYP is clearly less accurate than wavefunction schemes, though the MAE remains in line with other TD-DFT works. \cite{Lau13} Two simplifications were tested as well: 
i) removing the diffuse functions for the calculation of the adiabatic energies, which yields slight increases of the MSE by ca.~$0.04$ eV, but has rather negligible effects on the MAE; ii) using 
a $\Ezpve$ term obtained at the B3LYP/def2-TZVP level, which only yields a degradation of the MAE by ca.~$0.02$ eV, confirming the previously reported conclusion that this term can be safely estimated with a lower 
level of theory. \cite{Win13} In 2016, Oruganti, Fang, and Bo Durbeej \cite{Oru16} consider the WGLH set with the same philosophy as their 2014 work, \cite{Fan14b} i.e., finding simplified protocols delivering accurate 
0-0 energies. First, they showed that none of the tested XCF could deliver the same accuracy as CC2, the smallest MAE being obtained with B3LYP ($0.20$ eV), whereas, BP86 and M06-2X $\EOO$ deviate
much more significantly from experiment (MAE of $0.40$ and $0.36$ eV, respectively).  By using ZPVE corrections computed at the TD-DFT level, the changes on the CC2 $\EOO$ values are rather minor (roughly $0.04$ eV), whereas using 
CC2 for getting $\Ea$ and TD-DFT to determine both $\EreorgES$ and $\Ezpve$ led to variations ranging from $0.06$ to $0.12$ eV depending on the XCF, the hybrid functionals clearly outperforming BP86 (and CIS). 
\cite{Oru16} They concluded: \emph{``In fact, for a clear majority of the 66 states CC2-quality $\EOO$ can be calculated by employing CC2 only for the vertical term''}. The WGLH set was also chosen in 2017
by Schwabe and Goerigk in their investigation of spin-scaling effects on the transition energies obtained with double-hydrid XCFs. \cite{Sch17} Using the SCS-CC2 geometries of the original paper, they found that
both fitted and non-fitted variants of double hybrids behaved similarly. Using DSD-PBEP86/aug-cc-pVTZ to determine $\Ead$, they reached a MSE of $-0.02$ eV and a MAE of $0.06$ eV, \cite{Sch17} both values
being very similar to the one reported for the SCS-CC2 method. \cite{Win13}

In 2014, Barnes \emph{et al.} studied the $\EOO$ values determined for 29 transitions in 15 radicals (from diatomics to small aromatic systems). \cite{Bar14b} After having demonstrated that the 6-311++G(d,p) basis set
offered a good compromise, they investigated a wide range of XCF within the TD-DFT framework as well as CIS and CASPT2. While the usual CIS overestimation is extremely large (typically $> 1$ eV),
the performance of CASPT2 is quite remarkable with a MSE of $-0.02$ eV and a MAE of $0.12$ eV. At the TD-DFT level, the authors determined that the most valuable results are obtained with  
B3LYP, M06-2X, $\omega$B97X-D, and CAM-B3LYP for these open-shell systems. In contrast to other studies, no significant difference was noticed when separately considering the small
(di- and tri-atomics) and the medium-sized compounds.

In 2016, Tuna, Thiel and coworkers proposed an extended benchmark of their OMx/MRCI methods, including calculations of $\EOO$. \cite{Tun16} For 12 cases, they could compare the OM2/MRCI and B3LYP  
$\Ezpve$ and an average deviation of $0.04$ eV was found, a rather large value for this property, highlighting that the semi-empirical approach is not yet optimal to determine the ZPVE of ESs. As a consequence
they relied on TD-B3LYP $\Ezpve$ in their benchmark study. They investigated compounds of both Furche's 2011 and H\"attig's 2013 sets, discarding cases for which the OMx approaches were not 
parametrized. For the SKF set, the average errors are quite similar to TD-B3LYP (Table \ref{Table-1}), which is certainly a success.  However, the authors noted that OM2 and OM3 yield different error signs for the 
$\pi \rightarrow \pi^\star$ (underestimation) and  $n \rightarrow \pi^\star$ (overestimation) transitions, whereas TD-B3LYP consistently underestimate the 0-0 energies of both families of transitions. For the WGLH set, 
which is strongly dominated by $\pi \rightarrow \pi^\star$ transitions in aromatic organic molecules, the average errors are substantially larger with MAE of $0.35$ eV for both OM2/MRCI  and OM3/MRCI, and a 
clear trend to undershoot $\EOO$.

\begin{figure*}
  \includegraphics[width=\linewidth]{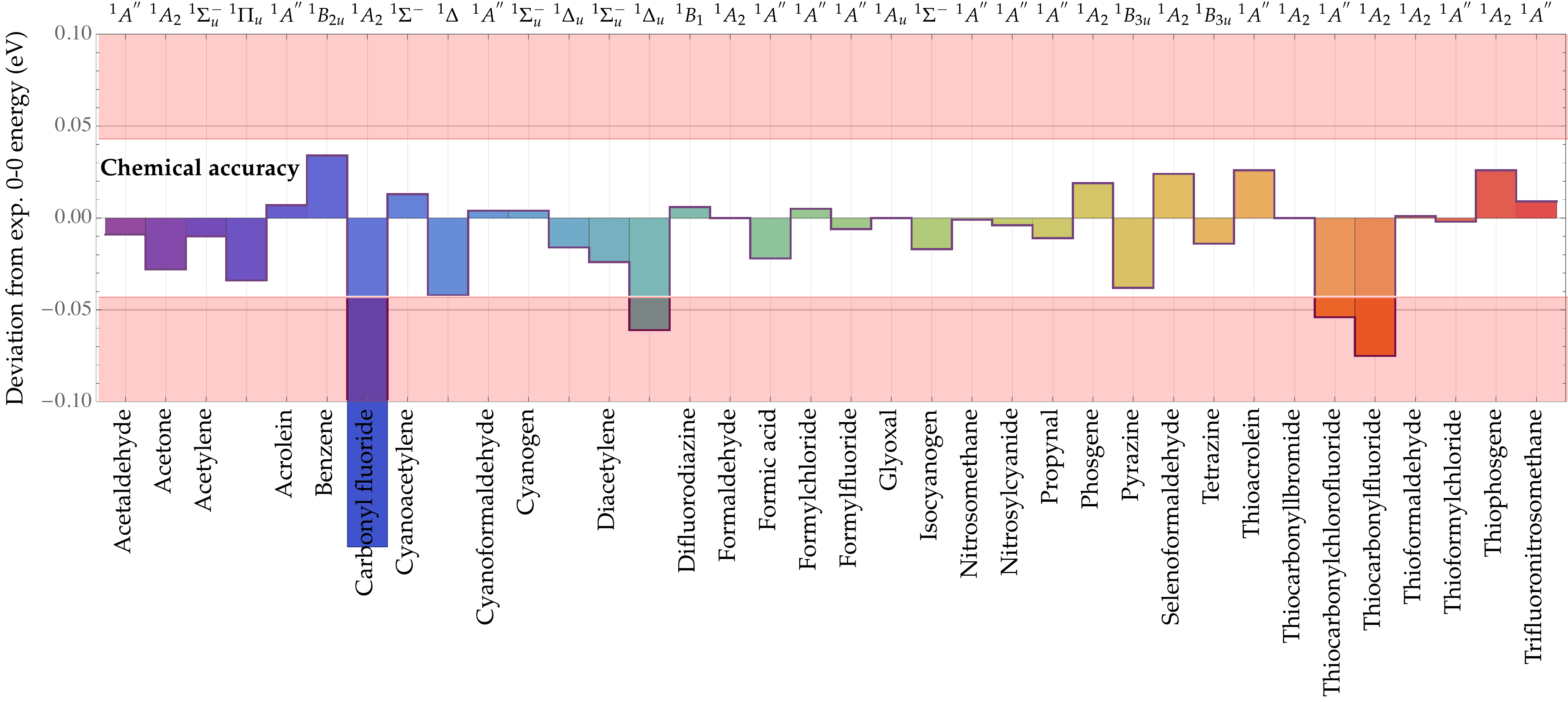}
  \caption{Deviation (in eV) from the experimental $\EOO$ of the theoretical $\EOO$ determined at the CC3//CCSDR(3) level. 
  Reproduced from Ref.~\citenum{Loo18b} with permission of the American Chemical Society.  Copyright 2018 American Chemical Society.}
   \label{Fig-3}
\end{figure*}

Recently, we have put some efforts in reaching very accurate $\EOO$ for non-trivial molecular systems. \cite{Loo18b,Loo19a} In our first contribution, we have considered singlet ES determined on molecules containing between 4 and 12 atoms
for a set encompassing more $n \rightarrow \pi^\star$ (25) than  $\pi \rightarrow \pi^\star$ (10) transitions. Using CC3 $\Ead$, CCSDR(3) geometries, and B3LYP $\Ezpve$, not only is the MAE very small ($0.02$ eV),
but chemical accuracy is achieved on an almost systematic basis (ca.~90\%\ success rate). The results for this set are illustrated in Figure \ref{Fig-3}. As one can be see, carbonylfluoride yields a significant deviation 
($-0.18$ eV), but it has been determined that this case is an outlier,  to be removed from the statistics, at the 99\%\ confidence level according to a Dixon $Q$-test. \cite{Loo18b} Data 
from Table \ref{Table-1} clearly demonstrate that using lower levels of theory than CC3 to determine $\Ead$ significantly degrades the results with MAE of $0.05$, $0.21$, and $0.08$ eV with CCSDR(3), CCSD and CC2, respectively.  Interestingly,
the CC2 MAE is similar to the one obtained on the WGLH set, whereas CCSD tends to exaggerate the transition energies, an observation consistent with other works. \cite{Sch08,Loo18a} In addition, using a quadruple-$\zeta$ 
basis set or including anharmonic corrections in the $\Ezpve$ term yield trifling variations for the data of Figure \ref{Fig-3}. \cite{Loo18b}

\begin{figure*}
  \includegraphics[width=\linewidth]{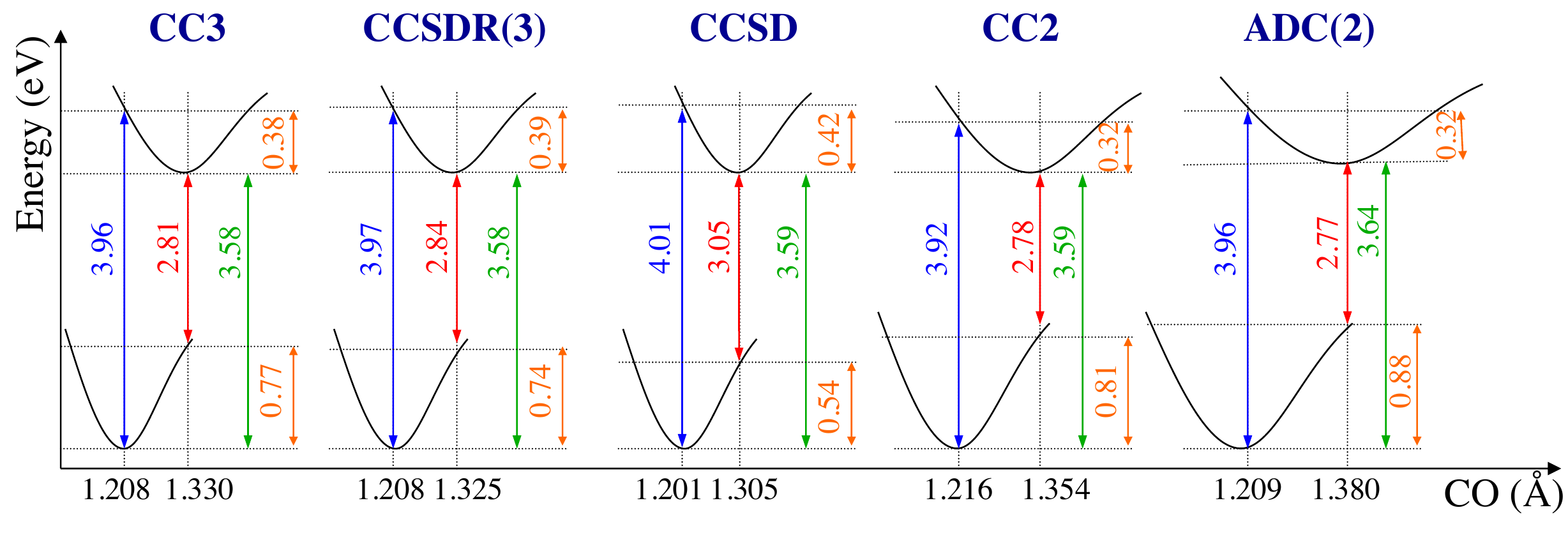}
  \caption{CC3/\emph{aug}-cc-pVTZ transition energies computed for formaldehyde computed with, from left to right, the CC3, CCSDR(3), CCSD, CC2 and ADC(2) geometries. 
  The absorption, fluorescence, adiabatic and reorganization energies are represented in blue, red, green and orange, respectively.  On the horizontal axis, we provide the optimal \ce{C=O} 
  bond lengths for these five geometries.  Reproduced from Ref.~\citenum{Loo19a} with permission of the American Chemical Society.  Copyright 2019 American Chemical Society.}
   \label{Fig-4}
\end{figure*}

In our most recent work, we have significantly increased both the size and the variety of the considered transitions (69 singlet, 30 triplet, 20 open-shell) with a focus set on the impact of the geometries 
on the computed $\EOO$. \cite{Loo19a}  First, the CC3 vertical and adiabatic energies determined on CC3, CCSDR(3), CCSD, CC2 and ADC(2) structures have been compared to a set of 31 singlet
transitions. Interestingly, while the level of theory considered to optimize the GS and ES geometries has a very strong impact on the vertical values, it has a very small influence on the adiabatic energies. 
For instance, taking the CC3//CC3 values as references, the MAE obtained with the CC3//CCSD method is $0.07$ eV for $\Ea$, $0.17$ eV for $\Ef$ but $0.01$ eV for $\Ead$.  Therefore, there is a clear 
error compensation mechanism taking place between the vertical and the reorganization energies, in the following expression
\begin{equation}
	\Ead = \frac{\Ea + \Ef}{2} +  \frac{\EreorgGS - \EreorgES}{2}. 	
\end{equation}
This has been illustrated for the case of formaldehyde (see Figure \ref{Fig-4}). On the CC3 geometry, $\Ead = 3.580$ eV, a value dominated by the first term of the previous equation (3.385 eV), the
second contributing to +0.195 eV. When going to other geometry optimization schemes, one notes significant changes of both terms with values $3.385$, $3.405$, $3.533$, $3.350$, and $3.364$ eV 
for the first, and $0.195$, $0.175$, $0.057$, $0.244$, and $0.278$ eV for the latter when using CC3, CCSDR(3), CCSD, CC2, and ADC(2) geometries, respectively. Nevertheless, their sum ($\Ead$)
is remarkably stable as seen in Figure \ref{Fig-4}. In addition, by comparing the experimental and theoretical 0-0 energies produced by combining i) CC3 $\Ea$, ii) CCSD geometries, and iii) B3LYP $\Ezpve$ 
corrections, a trifling MSE of $-0.01$ eV and a MAE of $0.03$ eV are obtained for the set of 119 transitions considered. \cite{Loo19a} Concomitantly, this means that, if $\Ead$ is determined at a 
high level of theory, one can obtain very accurate $\EOO$ even on geometries that cannot be considered as highly accurate. This could explain why some of the previous works \cite{Die04,Sen11b,Win13} 
noted small statistical fluctuations when going from, e.g., CC2 to B3LYP geometries. 

\section{0-0 energies in solution}

Performing comparisons between theoretical and experimental $\EOO$ energies determined in solution allows to tackle large compounds for which gas-phase measurements are beyond reach,
but obviously entails further approximations on the modeling side to account for environmental effects. In solution, experimental $\EOO$  values are generally taken as the absorption-fluorescence 
crossing point (AFCP) or the foot of the absorption spectra. The second choice is a cruder approximation in most cases, while the former limits the reference data to fluorescent compounds, that is, 
rather rigid derivatives. As noticed below, most published benchmark works use the polarizable continuum model (PCM) to describe solvation effects, \cite{Tom05} applying either its 
linear-response (LR), \cite{Cam99b,Cos01} its corrected linear-response (cLR), \cite{Car06} or its Improta's state-specifc (IBSF from the authors's name)\cite{Imp06} forms. 
The results obtained in published benchmarks are summarized in Table \ref{Table-2}.

%
%
\begin{longtable*}{lccllldd}
\caption{
\label{Table-2}
Statistical analysis of the results obtained in various benchmarks comparing $\EOO$ computations in solution to experimental data (AFCP). See caption of Table \ref{Table-1} for more details.
}
\\
\hline
\hline
Ref.				&	Year		& No.~of ESs	& No.~of molecules		& Method 			& Solvent	&	\mcc{MSE}		&	\mcc{MAE}		\\
\hline
\endfirsthead
\hline
\hline
Ref.				&	Year		& No.~of ESs	& No.~of molecules		& Method 			& Solvent	&	\mcc{MSE}		&	\mcc{MAE}		\\
\hline
\endhead
\hline \multicolumn{8}{r}{Continued on next page} \\
\endfoot
\hline
\hline
 \multicolumn{8}{l}{$^a${Extends the previous work by the same group, \cite{Goe09} see the text for details of the procedure;}}\\
 \multicolumn{8}{l}{$^b${(Sub)set of the JPAM set proposed in Ref.~\citenum{Jac12d};}}\\
 \multicolumn{8}{l}{$^c${Structures and ZPVE obtained in gas-phase with M06-2X/6-31+G(d);}}\\
\endlastfoot
\citenum{Goe10a}$^a$&	2010		& 	12		&12 (organic dyes)		& CIS/def2-TZVPP//PBE/TZVP			& LR-PCM	&	0.77		& 0.77	\\
				&			&			&					& CIS(D)/def2-TZVPP//PBE/TZVP		& LR-PCM	&	0.25		& 0.25	\\
				&			&			&					& SCS-CIS(D)'/def2-TZVPP//PBE/TZVP	& LR-PCM	&	0.33		& 0.33	\\
				&			&			&		& SCS-CIS(D)$^{\lambda=0}$/def2-TZVPP//PBE/TZVP	& LR-PCM	&	0.13		& 0.20	\\
				&			&			&		& SCS-CIS(D)$^{\lambda=1}$/def2-TZVPP//PBE/TZVP	& LR-PCM	&	0.03		& 0.19	\\
				&			&			&					& SOS-CIS(D)/def2-TZVPP//PBE/TZVP	& LR-PCM	&	0.07		& 0.19	\\
				&			&			&					&  CC2/def2-TZVPP//PBE/TZVP		& LR-PCM	&	0.00		& 0.17	\\
				&			&			&					&  SCS-CC2/def2-TZVPP//PBE/TZVP	& LR-PCM	&	0.15		& 0.20	\\
				&			&			&					&  BLYP/def2-TZVPP//PBE/TZVP		& LR-PCM	&	-0.49		& 0.51	\\
				&			&			&					&  B3LYP/def2-TZVPP//PBE/TZVP		& LR-PCM	&	-0.22		& 0.31	\\
				&			&			&					&  PBE38/def2-TZVPP//PBE/TZVP		& LR-PCM	&	0.04		& 0.19	\\
				&			&			&					&  BMK/def2-TZVPP//PBE/TZVP		& LR-PCM	&	0.07		& 0.19	\\
				&			&			&				&  CAM-B3LYP/def2-TZVPP//PBE/TZVP		& LR-PCM	&	0.11		& 0.18	\\
				&			&			&				&  B2PLYP/def2-TZVPP//PBE/TZVP			& LR-PCM	&	-0.11		& 0.20	\\
				&			&			&				&  B2GPLYP/def2-TZVPP//PBE/TZVP		& LR-PCM	&	-0.01		& 0.16	\\
\citenum{Jac12d}	&	2012		&	40		& 40 (organic dyes)	& B3LYP/6-311++G(2df,2p)//6-31+G(d)		& cLR-PCM	&	-0.14		&0.27	\\
				&			&			&				& PBE0/6-311++G(2df,2p)//6-31+G(d)		& cLR-PCM	&	-0.03		&0.22	\\
				&			&			&				& M06/6-311++G(2df,2p)//6-31+G(d)			& cLR-PCM	&	0.05		&0.23	\\
				&			&			&				& M06-2X/6-311++G(2df,2p)//6-31+G(d)		& cLR-PCM	&	-0.25		&0.26	\\
				&			&			&		& CAM-B3LYP/6-311++G(2df,2p)//6-31+G(d)			& cLR-PCM	&	-0.24		&0.25	\\
				&			&			&			& LC-PBE/6-311++G(2df,2p)//6-31+G(d)			& cLR-PCM	&	-0.56		&0.57	\\
\citenum{Cha13c}$^b$&	2013		&	7		& 7 (organic dyes)	& CIS/6-31+G(d)						& IBSF-PCM	&	0.75		&0.75	\\
				&			&			&				& TD-HF/6-31+G(d)						& IBSF-PCM	&	0.43		&0.43	\\
				&			&			&				& B3LYP/6-31+G(d)						& IBSF-PCM	&	-0.26		&0.30	\\
				&			&			&				& TDA-B3LYP/6-31+G(d)					& IBSF-PCM	&	-0.04		& 0.13	\\
\citenum{Jac14a}$^b$&	2014		&	40		& 40 (organic dyes)	&SOGGA11-X/6-311++G(2df,2p)//6-31+G(d)	& cLR-PCM	&	0.21		&0.24	\\
				&			&			&		&$\omega$B97X-D/6-311++G(2df,2p)//6-31+G(d)		& cLR-PCM	&	 0.30		&0.30	\\
				&			&			&			& LC-PBE*/6-311++G(2df,2p)//6-31+G(d)			& cLR-PCM	&	 0.12     	& 0.20 	\\
\citenum{Moo14}$^b$&	2014		&	40		& 40 (organic dyes)	& APD-D/6-311++G(2df,2p)//6-31+G(d)		& cLR-PCM	&	-0.06		&0.27	\\
				&			&			&				& PBE0-1/3/6-311++G(2df,2p)//6-31+G(d)		& cLR-PCM	&	 0.14		&0.22	\\
				&			&			&				& LC-PBE0*/6-311++G(2df,2p)//6-31+G(d)	& cLR-PCM	&	 0.25     	& 0.26 	\\
\citenum{Jac15b}$^c$&	2015		&	80		& 80 (organic dyes)	& M06-2X/6-311++G(2df,2p)//6-31+G(d)		& LR-PCM	&  	0.06		& 0.17	\\
				&			&			&				&									& cLR-PCM	& 	0.22		& 0.23	\\
				&			&			&				& CIS(D)/aug-cc-pVTZ//M06-2X/6-31+G(d)	& LR-PCM	&  	0.09		& 0.18	\\
				&			&			&				&									& cLR-PCM	& 	0.25		& 0.26	\\				
				&			&			&				& ADC(2)/aug-cc-pVTZ//M06-2X/6-31+G(d)	& LR-PCM	& 	-0.19		& 0.22	\\
				&			&			&				&									& cLR-PCM	& 	-0.03		& 0.14	\\				
				&			&			&				& CC2/aug-cc-pVTZ//M06-2X/6-31+G(d)		& LR-PCM	& 	-0.13		& 0.16	\\
				&			&			&				&									& cLR-PCM	& 	0.03		& 0.13	\\				
				&			&			&				& SCS-CC2/aug-cc-pVTZ//M06-2X/6-31+G(d)	& LR-PCM	& 	0.04		& 0.11	\\
				&			&			&				&									& cLR-PCM	& 	0.20		& 0.20	\\				
				&			&			&				& SOS-CC2/aug-cc-pVTZ//M06-2X/6-31+G(d)	& LR-PCM	& 	0.13		& 0.16	\\
				&			&			&				&									& cLR-PCM	& 	0.28		& 0.28	\\				
				&			&			&				& BSE/ev$GW$/aug-cc-pVTZ//M06-2X/6-31+G(d)& LR-PCM	& 	-0.14		& 0.19	\\
				&			&			&				&									& cLR-PCM	& 	0.02		& 0.15	\\				

\end{longtable*}

As stated in the previous Section, in their 2004 investigation Dierksen and Grimme applied an empirical correction to the experimental $\EOO$ measured in solution to obtain gas-phase reference values. \cite{Die04b}
In two more recent investigations, the same group proposed to transform experimental AFCP into solvent-free vertical estimates for, first, five \cite{Goe09} and, next, twelve \cite{Goe10a} dyes, by applying a series of additive 
theoretical corrections to the measured AFCP energies: i) solvation effects on $\Ea$ are determined at the LR-PCM/PBE0/6-31G(d) level, ii) zero-point vibrational corrections ($\Ezpve$) are computed at the 
PBE/TZVP level, and iii) reorganization effects (the difference between $\Ea$ and $\Ead$) are calculated at the same PBE/TZVP level. Such procedure allows to benchmark many levels of theory, as one only 
needs to compute gas-phase $\Ea$.  In this way, Goerigk and Grimme could obtain a MAE in the $0.16$--$0.20$ eV range for many approaches (see Table \ref{Table-2}), \cite{Goe10a} including CC2, several 
spin-scaled versions of CIS(D), two double-hydrid functionals (B2PLYP and B2GPLYP), as well as some hybrid functionals (BMK, PBE38 and CAM-B3LYP).  In contrast to the results obtained for the WGLH set, \cite{Win13}
both CC2 and SCS-CC2 do not significantly outclass TD-DFT in the Goerigk-Grimme set. It is unclear if this unusual observation originates from the nature of the molecules included in their set or the theoretical protocol itself.

In 2012, another set of 40 medium and large fluorophores was developed (JPAM set), \cite{Jac12d} and TD-DFT calculations of $\EOO$ were performed with a series of global and range-separated hybrid functionals using a 
fully coherent approach, i.e., the structures and ZPVE were consistently obtained for each functional used to compute $\Ead$. In Ref.~\citenum{Jac12d}, the authors note that there is an inherent difficulty when accounting 
explicitly for solvation effects during the calculations. Indeed, while $\Ead$ and $\EOO$ are equilibrium properties as they correspond to minimum-to-minimum energy differences, the absorption and fluorescence transitions 
are very fast processes and, in terms of solvation effects, should be viewed as non-equilibrium processes, meaning that only the solvent's electrons have time to adapt to the solute electron density's changes. \cite{Cam99b,Tom05}
Consistently, the AFCP is a non-equilibrium property as well. To resolve this apparent contradiction, an extra correction needs to be applied to the threoretical $\EOO$ values in order to allow a fairer comparison with experimental AFCP values.  
Using this protocol, a series of twelve hybrid functionals have been tested over the years on the JPAM set,  \cite{Jac12d,Jac14a,Moo14} including optimally-tuned \cite{Bae10b,Aut14a} versions of PBE (LC-PBE*) and PBE0 (LC-PBE0*). 
As can be deduced from Table \ref{Table-2}, the majority of the functionals lead to MAE in the $0.2$--$0.3$ eV range, the smallest deviations being obtained with PBE0 ($0.22$ eV) and LC-PBE* ($0.20$ eV).  The functionals including a rather 
large amount of \emph{exact} exchange, e.g., M06-2X and  CAM-B3LYP, significantly overestimate the experimental values, but they provide more consistent (in terms of correlation with experiment) AFCP energies than ``standard'' 
hybrid functionals like B3LYP and PBE0.  The LC-PBE* functional allows to obtain both a small MAE and a high correlation, but at the cost of tuning the range separation parameter for each compound. \cite{Jac14a} 
Consistently with the gas phase results discussed above, it was also shown that the band shapes are rather insensitive to the selected functional, \cite{Moo14} so that the choice of the functional can be driven
by the accuracy in modeling $\EOO$. A subset of the JPAM set was also used in 2013 in a comparison between TDA and TD-DFT $\EOO$ and band shapes. \cite{Cha13c}  With the B3LYP functional, the results were found to be 
substantially improved with TDA, but the authors warned that \emph{``using other exchange-correlation functionals might well lead to larger theory-experiment deviations with TDA than TD-DFT.'' }

\begin{figure*}
  \includegraphics[width=0.8\linewidth]{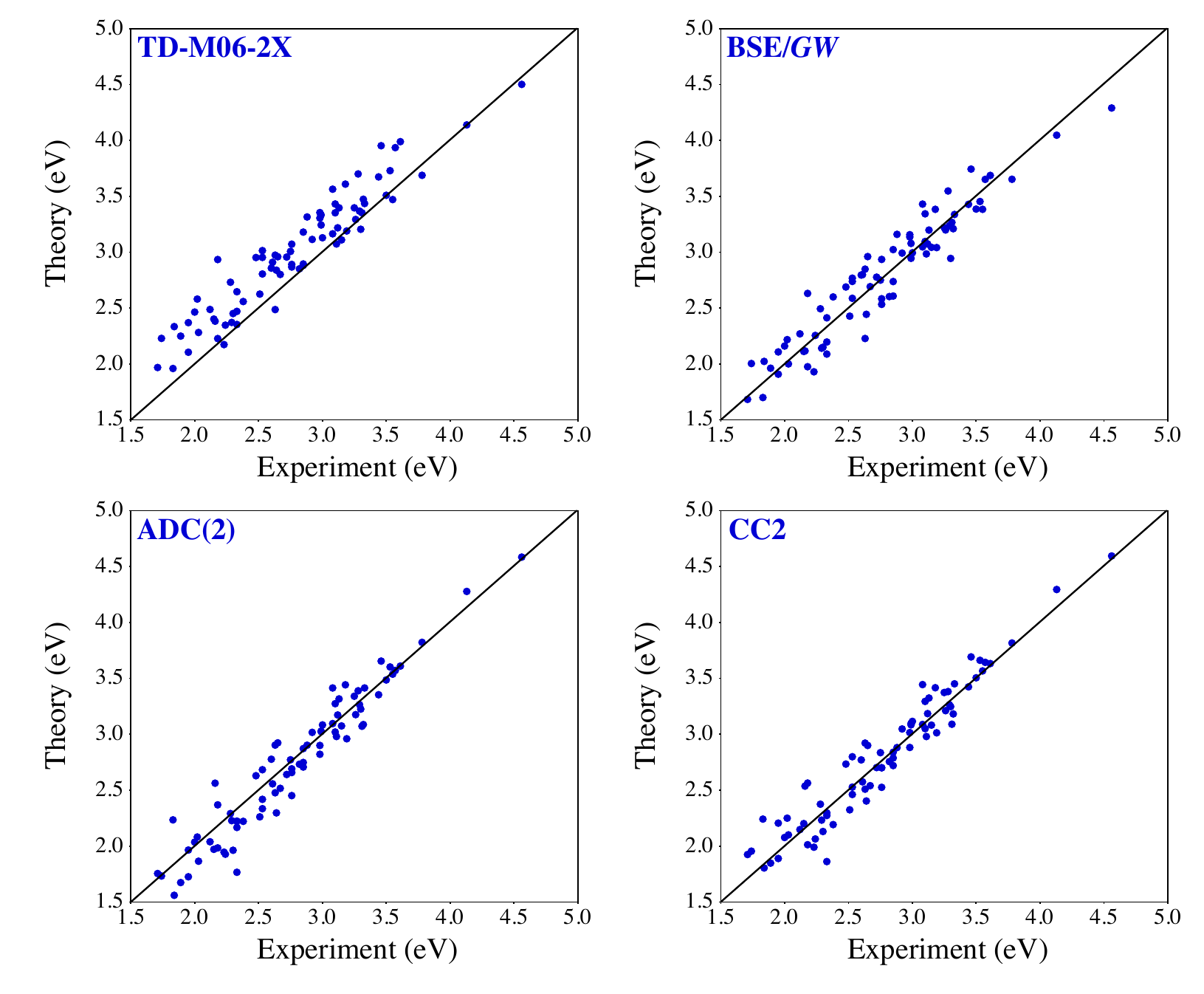}
  \caption{Correlation plots between experimental AFCP energies and theoretical $\EOO$ obtained for the JDB set applying the cLR-PCM solvent model. The central line indicates a perfect theory-experiment match. 
  Adapted from Figure 6 of Ref.~\citenum{Jac15b} with permission of the American Chemical Society. Copyright 2015 American Chemical Society.}
   \label{Fig-5}
\end{figure*}

\begin{figure*}
  \includegraphics[width=\linewidth]{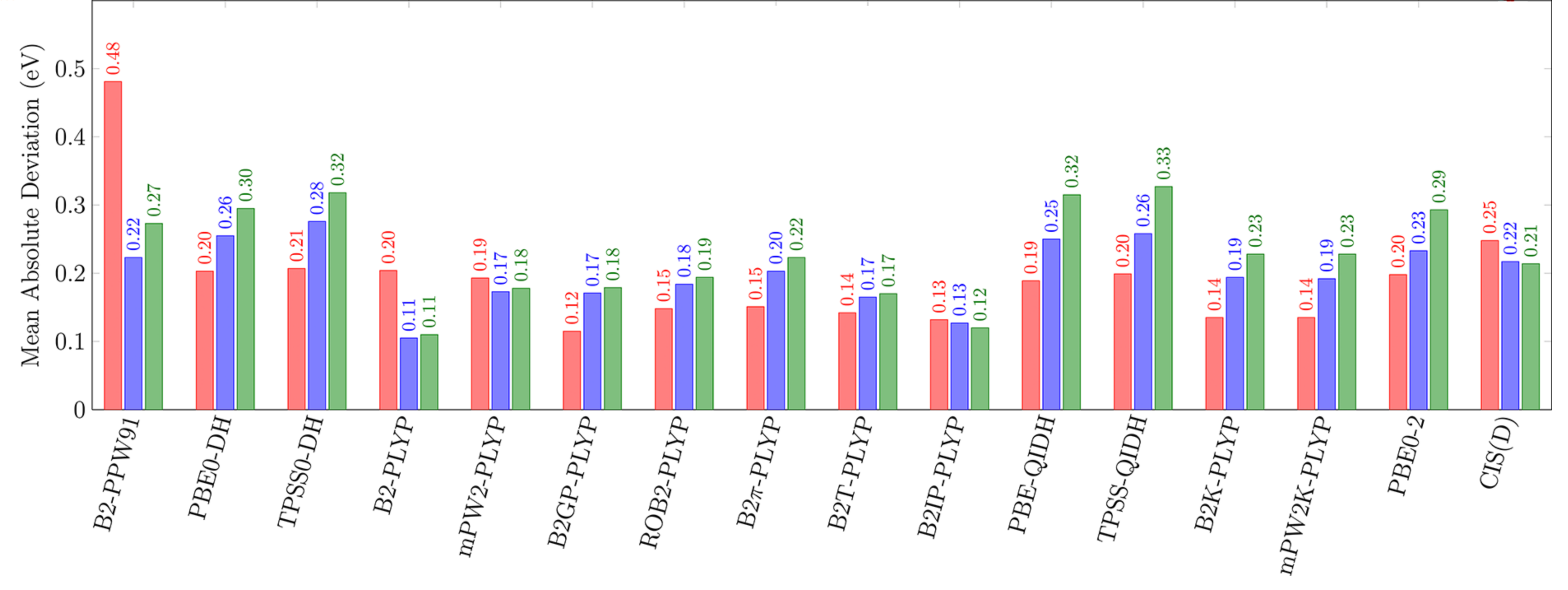}
  \caption{MAE (in eV) for $\Ea$ (red), $\Ef$ (blue) and $\Ead$ (green) computed with double-hybrid functionals for the EX7-1 subset of the JDB set using CC2 results as references.
  Reproduced from Figure 10 of Ref.~\citenum{Bre17} with permission of the American Chemical Society.  Copyright 2017 American Chemical Society.}
   \label{Fig-6}
\end{figure*}

In 2015, an even more extended set of fluorescent compounds (JDB set) was assessed using a protocol in which i) the structural and vibrational parameters are determined in gas phase at the M06-2X/6-31+G(d) level,
ii) the solvation effects are calculated as the difference of $\Ead$ computed in gas phase and in solution using LR-PCM or cLR-PCM, and iii) gas-phase $\Ead$ are determined using several wavefunction approaches in combination with the 
\emph{aug}-cc-pVTZ atomic basis set. \cite{Jac15b} As can be seen in Table \ref{Table-2} the selected solvent model has a large impact on the statistics, the LR-PCM $\EOO$ energies being almost systematically smaller 
than their cLR-PCM counterparts. \cite{Jac15b}  With the latter solvent model, the MAE are $0.13$, $0.14$, $0.15$, and $0.24$ eV with CC2, ADC(2), BSE/ev$GW$, and TD-M06-2X, respectively, the two former wavefunction methods 
providing higher determination coefficients as compared to experiment, as illustrated in Figure \ref{Fig-5}. \cite{Jac15b}  Given that the CC2 MAE obtained in gas phase on accurate geometries tend to be smaller (0.08 eV in 
Ref. \citenum{Loo18b}, $0.11$ eV in Ref.~\citenum{Oru16} and $0.07$ eV in Ref.~\citenum{Win13}), part of the $0.13$ eV error in this 80-compound set is probably due to the limits of the PCM models.  Consistently with the 
results obtained on the WGLH set, \cite{Win13,Oru16} the analysis of the data from the JDB set show that: i) ADC(2) and CC2 yield  very similar estimates, ii)  spin-scaling (SCS-CC2 and SOS-CC2) improves correlation with the 
experimental data but do not yield smaller MAE, and iii) the $\Ezpve$ term has a rather tight distributions around ca.~$-0.09$ eV.  With BSE/ev$GW$ the improvement with respect to TD-DFT is particularly significant for CT transitions, 
an expected trend for a theory explicitly accounting for the electron-hole interaction. \cite{Bla18} The $\Ea$, $\Ef$ and $\Ead$ data of the JDB set were also used by Adamo and coworkers to evaluate the performances of 
numerous double hybrid functionals. \cite{Bre16,Bre17}  In their second work, these authors found three subsets of the original JDB set able to reproduce the statistical errors of the complete set. Their most ``advanced'' subset (EX7-1) 
is composed of small molecules only, and therefore it allows rapid benchmarking as only computations on seven small compounds are needed to obtain relevant statistical results. Results obtained for the three families of transition energies with a wide range of 
double-hybrid functionals are given in Figure \ref{Fig-6}. Note that we did not included these results in Table \ref{Table-2} as Adamo and coworkers did not selected experimental data, but rather CC2 values, as references.

\section{Summary}

We have reviewed the generic benchmark studies devoted to adiabatic and 0-0 energies performed in the last two decades. Over the years, there has been a gradual shift from small to large molecules and from
gas-phase to solvents. Additionally, the level of theory has gradually increased. This can be illustrated by {the works benchmarking CC2}: whilst H\"attig's 2003 contribution was mainly 
devoted to di- and tri-atomics, \cite{Koh03} his group tackled much larger organic compounds only a decade later. \cite{Win13} Likewise, the first CC3 benchmark that appeared in 2005 only encompassed 19 states
in four diatomics, \cite{Hat05c} whereas more than 110 transitions in a diverse set of molecules (from 3 to 16 atoms) have been tackled recently. \cite{Loo19a} 

The results obtained in all these benchmarks, as measured by statistical deviations with respect to experimental measurements, are far from uniform, a logical consequence of the various protocols and molecular sets 
considered over the years. Nevertheless, some generic conclusions can be drawn:
\begin{enumerate}
\item  It is challenging to get a balanced description of various kinds of states ($n \rightarrow \pi^\star$ \emph{versus}  $\pi \rightarrow \pi^\star$, singlet-singlet  \emph{versus}  doublet-doublet...) and/or
  	  various families of compounds (small  \emph{versus}  large, organic   \emph{versus}  inorganic...). Therefore, we believe that benchmark's results focussing solely on a specific category of transitions/compounds
	  should not be generalized.
\item  In TD-DFT, for example, pure functionals, that do include \emph{exact} exchange, perform reasonably well for very compact compounds, but tend to provide significantly too low transition energies
	for medium and large derivatives, for which hybrid functionals have clearly the edge.
\item CC2 and ADC(2) yield similar accuracies, generally significantly outperforming CIS(D).  Globally, TD-DFT gives larger deviations than CC2 or ADC(2), except for double hybrids that are as accurate
         as these two approaches for a computational cost similar to CIS(D). These new functionals therefore represent a good compromise between accuracy and computational cost.
\item Spin-scaling approaches, e.g., SOS-CIS(D) and SCS-CC2, tend to provide more consistent data with respect to experiment but do not deliver smaller average deviations.
\item The total errors obtained for $\EOO$ are mainly driven by the errors on the transition energies, the level of theory used to obtain the structures having a rather minor impact on the results. 
         This outcome can be explained by an error compensation mechanism between the vertical and reorganization energies.
\item The $\Ezpve$ correction, the most costly contribution to 0-0 energies, is particularly insensitive to the methodological choice and is roughly equal to $-0.08$ eV for low-lying singlet-singlet transitions. One can therefore
   	select a low level of theory to compute it without significant loss of accuracy.
\item Given the two previous points, several simplified protocols can be used to compute more quickly $\EOO$. It is noteworthy that very compact test sets providing almost the same statistical values 
         have been developed recently.
\item The details of the approach employed to model solvation effects has a significant impact on the transition energies, hence, on the statistical results. At this stage, this conclusion holds for TD-DFT only,
  	as wavefunction-based benchmarks accounting for solvation effects have yet to appear.
\end{enumerate}

Given that calculations of  theoretical $\EOO$ offer well-grounded comparisons with highly refined experiments, the vast majority of the error comes from theory, and one can therefore provide a
rough estimate of the accuracy of various theoretical models, i.e., $1$ eV for CIS, $0.2$--$0.3$ eV for CIS(D), $0.2$--$0.4$ eV for TD-DFT when using hybrid functionals, $0.1$--$0.2$ eV for ADC(2) 
and CC2, and $0.04$ eV for CC3. Interestingly, rather similar error ranges have been obtained for CIS(D), ADC(2), CC2, and CC3, in recent comparisons with FCI data for small compounds, \cite{Loo18b} 
whereas the TD-DFT accuracy is globally the one found in comparisons with CC3 or CASPT2. \cite{Sil08}

\bibliography{CPC-00}

\end{document}